# Concept-Oriented Model and Query Language

Alexandr Savinov

SAP Research Center Dresden
Chemnitzer Str. 48, 01187 Dresden, Germany

**Abstract:** We describe a new approach to data modeling, called the concept-oriented model (COM), and a novel concept-oriented query language (COQL). The model is based on three principles: duality principle postulates that any element is a couple consisting of one identity and one entity, inclusion principle postulates that any element has a super-element, and order principle assumes that any element has a number of greater elements within a partially ordered set. Concept-oriented query language is based on a new data modeling construct, called concept, inclusion relation between concepts, and concept partial ordering in which greater concepts are represented by their field types. It is demonstrated how COM and COQL can be used to solve three general data modeling tasks: logical navigation, multidimensional analysis and inference. Logical navigation is based on two operations of projection and de-projection. Multidimensional analysis uses product operation for producing a cube from level concepts chosen along the chosen dimension paths. Inference is defined as a two-step procedure where input constraints are first propagated downwards using de-projection and then the constrained result is propagated upwards using projection.

# 1     Introduction

A model can be viewed as a mathematical description of a world aspect (Kaburlasos, 2006). One of the main goals of a data model is to provide means for organization of data that makes access to and manipulations with data easier for users. Organization of data consists in using some structures which allow us to break the whole space of data elements into smaller groups where data can be accessed and manipulated more productively. Currently, there exist many such structural principles through which data space can be viewed such as relation, hierarchy, graph or multidimensional cube. This variety is due to the fact that "no single model of reality may be appropriate for all users and problem domains … and the mechanisms by which they are most naturally referenced vary across different world views" (Shipman, 1981). In this paper we present a new alternative approach to data modeling, called the concept-oriented model (COM), which is based three structural principles. First, any data element is supposed to consist of two parts: identity and entity. Second, all elements exist within inclusion hierarchy. And third, all elements are partially ordered. To describe the model we define a novel query language the syntax of which reflects these principles.

What are the major motivations behind COM? The first motivation is based on one obvious but very important empirical observation that everything needs to be somehow *represented* before it can be accessed, characterized or otherwise manipulated, and the properties of this representation, called *identity*, differ significantly from the properties of the represented thing, called *entity*. The role of identities has never been underestimated and they have been extensively investigated in data modeling, programming languages and other fields of computer science. There exist numerous studies (Khoshafian et al, 1986; Wieringa et al, 1995; Kent, 1978; Abiteboul et al, 1989; Abiteboul et al, 1998; Kent, 1991; Eliassen et al, 1991) highlighting them as an essential part of database systems and demonstrating the need to have a strong and consistent notion of identity in data and programming models. A lot of approaches have been proposed for describing identities such as primary keys (Codd, 1970) in relational model, object identifiers (Abiteboul et al, 1989; Abiteboul et al, 1998; Khoshafian et al, 1986) in object-oriented database systems, l-values (Kuper et al, 1984, 1993), surrogates (Abrial, 1974; Hall et al, 1976; Kent, 1978; Codd, 1979) and many others. However, the problem is that the main focus in data modeling has been made on entities while identities are still treated as second-class elements. A typical data model provides only platform-specific identities while domain-specific identities are assumed to be modeled via entities without direct support from the model or system.

The main goal of COM in this context is to make identities at least as important as entities so that both are equal elements of the data model. Both identities and entities should have domain-specific structure and have to be modeled using dedicated means. To reach this goal COM takes a new approach by postulating in its duality principle that identity and entity are considered two constituents within one element rather than separate elements. Thus data modeling in COM is broken into two symmetric branches of identity and entity modeling by producing a nice yin-yang style of balance between two sides of one reality. Data modeling is then reduced to modeling such identity-entity pairs in their inseparable unity where identities are completely legalized by getting equal rights with entities. Very informally, it could be compared with the introduction of complex numbers in mathematics which also have two constituents – a real part and an imaginary part – manipulated as one whole. The effect is the same as in mathematics: this makes data modeling simpler and more natural. The main general purpose of identities in this design is to manifest the fact of *existence* and duality principle answers the question *how* elements exist.

Another motivation behind COM is based on the observation that any identity exists in some space, also called domain, context or scope. This means that a thing is not able to exist *in vacuo* by itself without anything outside of it: a thing with no context or a thing outside any space is nonsense. Therefore, the notion of space is as important as the identity itself because without the domain (context) we are not able to use its identities at all so it is one of the crucial concerns in identity modeling (Kent, 1979). Further, a context may have its own context so we get a hierarchical structure. Similar structure is the basis of the hierarchical model of data (Tsichritzis et al, 1976) and it underlies inheritance relation in object-oriented models (Dittrich, 1986) where it is used for modeling types (Albano, 1983). The notion of context and scope also exists in many conceptual approaches like that of Kent (1991) where scope is a special element of the model.

To provide a mechanism for modeling spaces, COM postulates in its inclusion principle that all elements exist within a hierarchy where each of them has one parent element and many child elements. The main purpose of this hierarchy is to manifest that elements exist in space and always have a context, domain or context represented by the parent element. Thus inclusion principle answers the question *where* elements exist. Another important role of this hierarchy is that it provides an address space where any element is uniquely identified by a sequence of relative identities starting from the root element and ending with this element. Such a sequence of simple or local identities is called a complex identity also known as a hierarchical address, compound identifier (Kent, 1979) or layered reference. Although employing hierarchies for data modeling is definitely not new, the distinguishing feature of COM is their interpretation, roles and uses: it provides hierarchical identities, it generalizes inheritance and it turns any element into a set which allows us to treat COM as a set-based model where any element is intrinsically a set.

The third motivation behind COM is data semantics and the goal here is to provide richer high-level mechanisms and constructs for representing complex application-specific concepts and relationships. Just as with identities and hierarchies, there has been a tremendous interest in semantic models and there exist numerous works devoted to investigating what data semantics is and how it can be modeled (Tsichritzis et al, 1982; Hull et al, 1987; Peckham et al, 1988; Potter et al, 1988). A variety of general modeling methodologies and specific techniques for the representation of data semantics have been proposed such as first binary models (Abrial, 1974), extensions to relational model (Codd, 1979) and sophisticated full-featured semantic modeling frameworks (Hammer et al, 1978; Kent, 1979; Hammer et al, 1981; Abiteboul et al, 1987; Jagannathan et al, 1988).

COM proposes a new view on data semantics by assuming in its order principle that semantics can be represented using partial order relation where any element has a number of greater and lesser elements. In this case the meaning of an element depends on its position among other elements in the partially ordered structure. Using partially ordered sets significantly simplifies data modeling because many mechanisms and patterns can be explained in terms of partial order relation. To represent a partial order COM employs references by assuming that a reference represents a greater element. It is one of the most important assumptions in COM because references change their role from navigational tool (in graphs) to an elementary semantic unit (in a partially ordered set).

Just as for any general approach which aims at changing the way how data is viewed and manipulated, it is highly difficult to provide a complete list of motivating factors that have driven this research and resulted in this model because they have frequently changed their form and substance. In addition, many of these motivating factors have inherently informal character and can hardly be formulated in a rational and logical way. For instance, the goal of the functional data model was to provide "naturalness and simplicity" (Peckham et al, 1988) and the goal of DAPLEX was formulated as providing a "conceptually natural" database interface language "which allows the user to more directly model the way he thinks about the problems he is trying to solve" (Shipman, 1981). COM shares these goals and also tries to reach a higher degree of simplicity and harmony. Another rather general motivating factor is that we would like to remove or at least to decrease the deeply root incongruity between data and programming models (Copeland et al, 1984; Atkinson et al, 1987). This is why COM has been developed as an extension of a novel approach to programming, called concept-oriented programming (COP) (Savinov, 2005b, 2008a, 2009b). At general level, COM can be viewed COP plus data semantics (partial ordering).

The main purpose of this paper is to describe COM using a novel language, called concept-oriented query language (COQL). The basic construct of COQL is concept (hence the name of this model). Concept is defined is a couple consisting of two classes: identity class and entity class. If identity class of concept is empty then we get a conventional class. So the difference is that if classes are used to model entities then concepts are used to model identity-entity couples. Inclusion relation between concepts generalizes inheritance which is used to define a hierarchical address space where data elements exist. Each concept declares its parent concept and then its instances will be identified relative to the parent instance. Partial order relation among concepts is established by its field types which specify greater concepts in the concept-oriented schema. Ordering concepts is also a new feature which has no close analogues in other data models. This principle leads to a new role of fields: now they represent greater elements rather than serve as storage for arbitrary links to other nodes in the graph. For data access, COQL defines two operations of projection and de-projection which rely on the partially ordered structure induced by field types. In summary, COM can be viewed as a syntactic embodiment of COM because it follows main patterns of thought employed in COM and implements its three main principles:

- Duality principle (how elements exist): an element has two constituents – identity and entity, the type of which is described by concepts

- Inclusion principle (where elements exist): an element has a parent element (interpreted as a space, domain, context or scope) where it exists which is declared by concept inclusion relation

- Order principle (what elements mean): an element is characterized by other elements defining its meaning which are specified in concept fields

Importantly, all the three principles (duality, inclusion and order) are logically connected because partial order needs references for its representation. References in turn need inclusion relation for modeling identities and hierarchical address spaces. And at the very basic level, we need duality to split identities and entities for modeling two branches separately.

Recently, a number of papers have been published which introduce and study various aspects of this emerging approach to data modeling taking into account different motivating factors and problems (Savinov, 2005a, 2006a, 2006b, 2009a). However, they describe either preliminary results or specific mechanisms and therefore the terms, definitions and problem formulations frequently change. In this article we summarize and generalize these results by unifying terminology used in COM/COQL. In particular, by concept we mean a language construct used as a type (the same as in COP) while in previous papers concept was defined as a collection. In this article we use prefixes super- and sub- only in the context of inclusion relation (again, to unify terminology with COP). To describe the same in the context of partially ordered structure we use terms greater and lesser (concept, element, collection etc.) In contrast, in previous papers prefixes super- and sub- were used in the context of both hierarchical and partially ordered structures. In diagrams, we use the convention that inclusion relation spreads horizontally while order relation spreads vertically. Fragments of queries which are being currently discussed are underlined.

The rest of the article has the following layout. Section 2 is devoted to describing three principles of the model: duality, inclusion and order. In particular, in this section we introduce the main data modeling construct, concept, define inclusion relation on concepts and describe how concepts are partially ordered in the concept-oriented schema. In Section 3 we describe various interpretations of order relation. Section 4 demonstrates how COM can be used for solving typical data modeling tasks: logical navigation, multidimensional analysis and inference. Section 5 is an overview of related work with discussion and Section 6 makes concluding remarks with an outlook to future research.

## 2 Principles of the Concept-Oriented Model

### 2.1 Duality Principle

In the concept-oriented model, a database is a set of *elements* or things. Duality principle postulates that an element has two constituents: one *identity* and one *entity*. Modeling separately identities and entities is a common practice in computer science but the distinguishing feature of COM is that an element is viewed as one whole. Strictly speaking, it is not possible to declare only an identity or only an entity but on the other hand they have completely different roles and properties within the element. In other words, we do not say that there exist entities and identities which can be somehow associated. Rather, it is postulated that there exists an element which has two sides or flavors, called identity and entity.

The main reason for distinguishing identity part and entity part within an element is that they have completely different properties and roles which are crucial for the whole model. Entity is regarded as a thing-in-itself which is radically unknowable reality not observable in its original form. In contrast, identity is a phenomenon observable and manipulated directly in its original form as it is. Entity represents a persistent state of an element while identity is its transient part. Entities are passed and stored by-reference while identities are passed by-value. Therefore, the state of an entity is located in one point and is shared among all its users while the state of an identity belongs to one user and hence no changes will be visible to others.

There are two major uses of identities: identity as a *reference*, and Identity as a *value*. If identity is interpreted as a reference then its purpose is to *represent* an entity. The entity represented by a reference is referred to as an *object*. References provide a mechanism for implementing links or connections in the model where we can store an identity (one constituent of an element) as a representative of the entity (the other constituent). Only a reference knows where the entity is located and how it can be accessed.

Importantly, both identity and entity have their own structure and behavior which have to be also modeled independently using appropriate techniques. But on the other hand, it is necessary to take into account that they constitute one whole. One of the contributions of COQL is that it introduces a novel data modeling construct, called *concept*, which is defined as follows:

> **Definition 1 [Concept].** *Concept* is a couple of two classes – one identity class and one entity class.

Concepts are intended to be a generalization and complete substitute for conventional classes. This means that if identity class of a concept is empty then this concept is equivalent to a class. In contrast to classes, a concept instance (element) is a couple consisting of one identity and one entity where identity part is stored in the variable. Concept fields are referred to as *dimensions* because this term emphasizes their semantic role (described in Section 2.3) which is absent in conventional classes. For example (Fig. 1), if banks are identified by their BIC (Bank Identifier Code) consisting of 11 characters while bank name and address are their entity (persistent) properties then the type of such elements is described by the following concept:

```
CONCEPT Bank                              // Concept name
  IDENTITY                                // Identity class
    CHAR(11) bic                   // Identity dimension
  ENTITY                                  // Entity class
    CHAR(64) name                  // Entity dimension
    Address address          // Stores identity of address
```

One consequence of introducing concepts is that the whole domain of data modeling is broken into two branches: *identity modeling* and *entity modeling*. They can be regarded as two orthogonal branches which are modeled in their unity by means of concepts. Although identity modeling by itself has always been known as an important task, only in COM it is made dual to and inseparable from entity modeling. Identity modeling is not only completely legalized within concepts (and via other mechanisms described further in the article) but these two branches are tightly related by naturally cross-cutting each other.

When concept is used as a type then variables, dimensions (fields), parameters and other typed references will store only identity of the referenced element with the user-defined structure. In contrast, if it were a conventional class then they would store a pointer, surrogate or other kind of primitive reference. In the above example, concept `Bank` has dimension `address` of type `Address`. Since `Address` is a concept, this dimension will contain an reference as defined in the identity class of the `Address` concept.

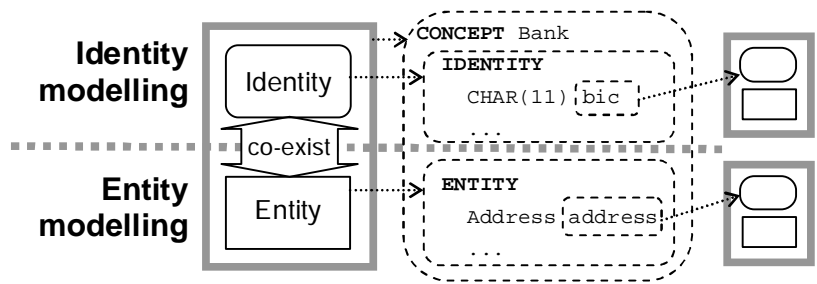

*Figure 1*. An element consists of one identity and one entity which is modeled by a concept.

In addition to using concepts as types of individual elements like dimensions of other concepts, their second use consists in specifying a type of elements within a collection. In this case two types are needed: one for the collection itself (like `Table` or `Array`) and the other for its elements (like

`Bank` or `Address`). The collection type is normally provided by DBMS while the type of its elements is specific to the problem domain and has to be provided by the designer as some concept name. For example, a table for storing bank records could be created using an SQL-like syntax as follows:

```
CREATE TABLE Banks CONCEPT Bank
```

This statement creates a new collection named `Banks` which is declared to contain only elements of concept `Bank` specified after the CONCEPT keyword. As a matter of convention, concept names will be always in singular while collections will have the same name but in plural. In the above example the concept has name `Bank` so the table with elements of this concept is named `Banks`.

A subset of elements can be selected from a collection by imposing simple constraints on its properties. The constraints will be written after the source collection name separated by a bar symbol. For example, if we need to select all banks with the name starting from 'A' then it is written as follows:

```
ResultCollection = ( Banks | name STARTSWITH 'A' )
```

In more verbose form the same statement will use an instance variable `b` which references the current element of this collection:

```
ResultCollection = ( Banks b | b.name STARTSWITH 'A' )
```

It is important to understand that COM collections are different from relational tables and this difference is discussed in Section 5.1 where we also discuss how COM identities are related to other existing identification mechanisms.

## 2.2 Inclusion Principle

Identity class can be used to describe an address space where one identity is one address. However, addresses within an address space always have a relative form because they are valid only within one domain. For example, a street name is specified with respect to some city and a bank account is meaningful only within its bank. Then the main question is what an address space is and what role it plays in the model? In COM, it is assumed that address spaces are normal elements with their own identity and entity constituents. Inclusion principle postulates that any element has a super-element and the identity of an element is specified relative to the identity of its super-element. To model this hierarchy at the level of concepts COM introduces a special facility, called *inclusion relation*:

> **Definition 2 [Concept inclusion]**. Concept has a super-concept specified via inclusion relation.

For example (Fig. 2), if a bank account is known to exist in the space of its bank and is identified relative to the bank then concept `Account` has to be included in concept `Bank` using keyword 'IN':

```
CONCEPT Account IN Bank              // Bank is super-concept
   IDENTITY
     CHAR(10) accNo
   ENTITY
     DOUBLE balance
     Person owner
```

Inclusion relation produces a hierarchy of concepts where any concept has one super-concept and may have many sub-concepts. The root of this concept hierarchy has a super-concept provided implicitly by DBMS (as a platform-specific type). Inclusion hierarchy will also be referred to as *physical* structure in order to distinguish it from the partially ordered structure described in the next section and called logical structure. Super- and sub-concepts within the physical hierarchy will also be referred to as (physical) parents and children, respectively. For example, if one account element is supposed to have internal sub-accounts then they can be described by sub-concepts `SavingsAccount` and `CheckingAccount` as shown in Fig. 2. Concept `Bank` is the root concept of this inclusion hierarchy.

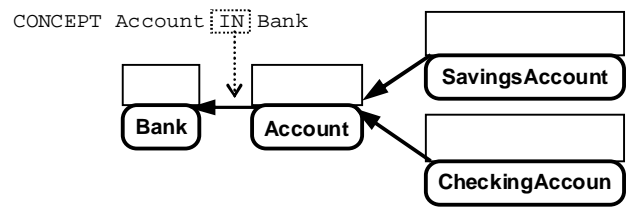

*Figure 2. Concept inclusion hierarchy.*

Parent elements in the inclusion hierarchy are interpreted as a (physical) context, domain or address space for their child elements. For example, a bank is an address space for its accounts which in turn are address spaces for internal sub-accounts. To fully identify an element within such a hierarchical address space, its identity has to contain all the parent identities up to the root.

> **Definition 3 [Complex identity].** *Complex identity* or complex reference is a sequence of simple identities, called identity segments, where concept of each next segment is included in the concept of the previous segment.

A sequence of entities represented by a complex identity is referred to as a *complex entity* or complex object. Complex identities are used to represent elements within a hierarchy where each identity segment is a local identity in the context of its parent element. For example, if concept `SavingsAccount` is included in concept `Account` which in turn is included in concept `Bank` (Fig. 3) then its instances will be identified by three segments: bank, account and sub-account. If we declare a variable of this type

```
    SavingsAccount savAcc;
```

then it will contain consecutively all three segments in the format of their identity classes defined in concepts `Bank`, `Account` and `SavingsAccount`. Apparently, inclusion relation can be viewed as a mechanism of identity extension or specialization where an included (child) concept adds more specific information to the parent identity which allows for representing more specific elements. If identity is a reference then we get more specific addresses and if it is a value then we get a more specific value (a value with additional fields in its structure).

Importantly, a dimension can store references of more specific sub-concepts than declared in its type (just like in object-oriented models). In other words, if a dimension is of some concept then it can store a reference to an element of this concept or any of its sub-concepts. This means that concepts provide a mechanism for specifying a container where referenced elements can reside. For example, if we declare a variable of type `Account`

```
    Account acc;
```

then it can reference any element which has some account as a parent. In particular, this variable can store a reference to a savings account of concept `SavingsAccount` because it is included in `Account`.

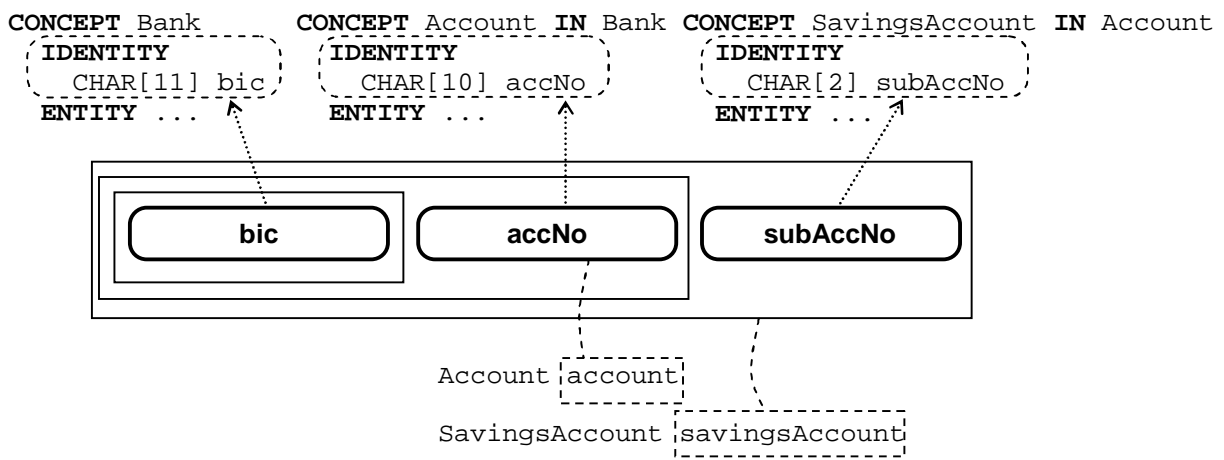

*Figure 3.* Structure of complex identity.

Just as concepts generalize conventional classes, concept inclusion generalizes class inheritance. Inclusion turns into inheritance if identity class is empty and hence instances of this concept cannot be distinguished in their domain. As a consequence, it is possible to create only one sub-element which is effectively treated as an extension inheriting properties of its super-element. For example, we could define a special savings account as an extension of already existing concept without identity class:

```
CONCEPT SpecialSavingsAccount IN SavingsAccount
  IDENTITY // Empty
  ENTITY
    INT privileges              // Extends its parent fields
```

This account has no identity and hence it is not possible to create many special accounts within its parent. Therefore, it will simply extend its parent precisely as it is done in object-oriented approaches. Essentially, we add one dimension in its entity (with special conditions for this account).

Inclusion also turns into inheritance if entity class is empty but in this case it is applied to values. For example, if we have a parent concept for describing figure coordinates then we could extend it by adding an additional dimension for describing its size:

```
CONCEPT Figure
  IDENTITY
    INT x, y
  ENTITY // Empty
CONCEPT FigureWithSize IN Figure
  IDENTITY
    INT size                    // Extends its parent fields
  ENTITY // Empty
```

An instance of concept `Figure` has two dimensions while an instance of concept `FigureWithSize` has three dimensions. Yet, they both describe values.

Generally, the main difference between inclusion and inheritance is that instances of concepts exist within a hierarchy while instances of classes exist in flat space. In other words, just as parent concepts may have many child concepts, parent elements may have many child elements. Using object-oriented terminology this means that a base object in COM may have many extensions and, vice versa, many extensions may share one base object. In contrast, any extension in OOP has its own base (from which it inherits its identity). Another difference between inclusion and inheritance is that entity segments are separate objects with their own identity and life-cycle. So an extension can be created or deleted within its base element. In OOP it is not possible because they have the same (primitive) identity.

Let us now consider how inclusion relation is used when defining collections. Any collection is supposed to store only instances of its concept while parent instances and child instances are stored in separate collections. Therefore, when a new collection is created it is necessary to specify a concrete parent collection of the parent concept and this procedure is referred to as *binding*. The declaration of a parent collection is analogous to declaration of a parent concept but the difference is that it is done for any new collection. For example, if we create a new table with accounts then we have to specify a table where its parent elements (banks) will be stored:

```
CREATE TABLE Banks CONCEPT Bank
CREATE TABLE Accounts CONCEPT Account IN Banks
```

Now the system knows about the hierarchical connections and can use them for maintaining, navigating through or querying the inclusion hierarchy. The keyword 'super' is used in queries to refer to the super-element. For example, in order to print accounts along with their bank name the following SQL-like query could be used:

```
SELECT acc.accNo, acc.super.name FROM Accounts acc
```

Here `acc` is an instance variable referencing the current account from table `Accounts` and `acc.super` references the bank segment of the current account from table `Banks`.

## 2.3  Order Principle

In the previous section we assumed that elements exist in an inclusion hierarchy which provides a hierarchical address space for them. However, identity modeling is only one branch in COM which is intended for describing how elements are represented and accessed. The second branch is entity modeling where elements use each other to describe their semantics. Here the main idea of COM is that all elements are partially ordered and inclusion principle postulates that any element has a number of *greater* elements which are represented by their identities stored in the dimensions of this element.

Essentially, we make two assumptions: (i) all elements are partially ordered and (ii) this order is represented by references stored in dimensions. The second assumption is also a very important distinguishing feature of COM because it states that a reference (a stored identity) represents a greater element in the partially ordered set (poset): if element *a* stores a reference to element *b* in its dimension *x* then *a* is a *lesser* element and *b* is a *greater* element, $a <_x b$. The lesser element *a* will also be called a *logical child* while the greater element *b* will be called a *logical parent*. Why ordering represents semantics? Because semantics of an element is its meaning stored in its definition and defining an element means reducing it to more general elements. In COM, an element is defined via its greater (referenced) elements which in turn are defined via their greater elements and all together they define semantics. Semantics also allows for reasoning about data and this mechanism in COM is based on partial order relation among its elements.

References in COM are not simply a means of connectivity within a graph or network. Rather, a reference is an elementary semantic construct because it reduces the element to a more general (referenced) logical parent which is a constituent in its definition. Storing a reference is an important step because it influences semantics of this element. For example, if element *a* references element *b* then what can be said about their meaning? In most existing models, its meaning is formally or informally encoded in the dimension description while in COM the formal meaning is that *b* is greater than *a* (other interpretations are described in Section 3). Thus the idea of COM is that references are a means for bringing semantics into a set of elements by establishing partial order and this order is then used for semantic operations including querying and inference.

Just as references are used to represent greater elements, dimension types are used to partially order concepts using the following definition:

> **Definition 4 [Concept ordering]** Concept has a number of greater concepts specified by its dimension types.

Any concept has as many greater concepts as it has dimensions (in both identity and entity classes). Conversely, each use of this concept as a dimension type within some other concept means the existence of one lesser concept (logical child). For example (Fig. 4), if concept `Person` has a dimension of concept `Address` then `Person` is a lesser concept and `Address` is a greater concept. And if concept `Bank` is also characterized by an address of concept `Address` then `Address` has two lesser concepts `Person` and `Bank`. Note that one greater concept may have many lesser concepts (like concept `Address` is a greater concept for both `Person` and `Bank`) and one lesser concept may have many greater concepts because it may have many dimensions.

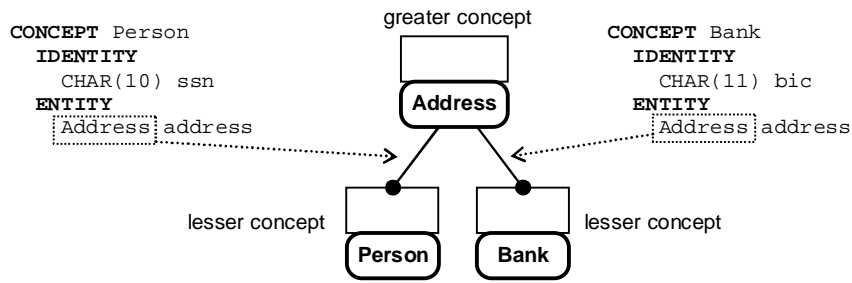

*Figure 4. Concept ordering by means of dimension types.*

Concepts without greater concepts are referred to as *primitive concepts*. Normally these are primitive (platform-specific) data types like integer but it can also be an arbitrary concept if we want to abstract from its structure of dimensions. Semantically, primitive concepts represent basic notions the meaning of all other notions in the model is reduced to. The greatest concept which is a direct or indirect parent of all other concepts is called *top concept*. It is introduced formally as empty concept and is a greater concept for primitive concepts. The least concept which is a direct or indirect child for all other concepts is referred to as *bottom concept*. Bottom concept can be introduced formally.

Concept-oriented schema in COM is defined using two relations: inclusion and partial order.

> **Definition 5 [Concept-oriented schema].** *Concept-oriented schema* is a number of concepts where each concept has one super-concept defined by inclusion relation and many greater concepts defined by its dimension types.

Any concept in a concept-oriented schema belongs to two structures simultaneously. Hierarchical structure induced by inclusion relation is used for identity modeling and describing how data is represented and accessed. The orthogonal partially ordered structure induced by dimension types is used for entity modeling and describing data semantics. In diagrams, we will assume that inclusion spreads horizontally while ordering spreads vertically. This means that all sub-concepts (physical children) are positioned to the right of their super-concept and all lesser concepts (logical children) are positioned under their greater concepts. Inclusion relation is denoted by leftward arrows pointing to a super-element. Ordering is denoted by line segments starting from a lesser concept (marked by a black circle depicting a dimension) and leading up to its greater concept.

An example of a concept-oriented schema consisting of 8 concepts is shown in Fig. 5. Here we assume that the first segment of any address is city while the second segment is street, house number etc. Therefore, concept `Address` is included in concept `City` (1). Both persons and banks are characterized by some address and therefore concept `Address` is a greater concept for concepts `Person` and `Bank` (2). This means that any person element and any bank element will have a dimension storing a complex reference of some address element. Accounts are defined in the context of their bank and hence concept `Account` is included in concept `Bank` (3). A main account may have more specific sub-accounts defined in its context via a child concept like `SavingsAccount` and `CheckingAccount` included in `Account` (4). We also assume that there is many-to-many relationship between accounts and their owners and therefore concept `AccountOwner` has two greater concepts: `Person` and `Bank` (5).

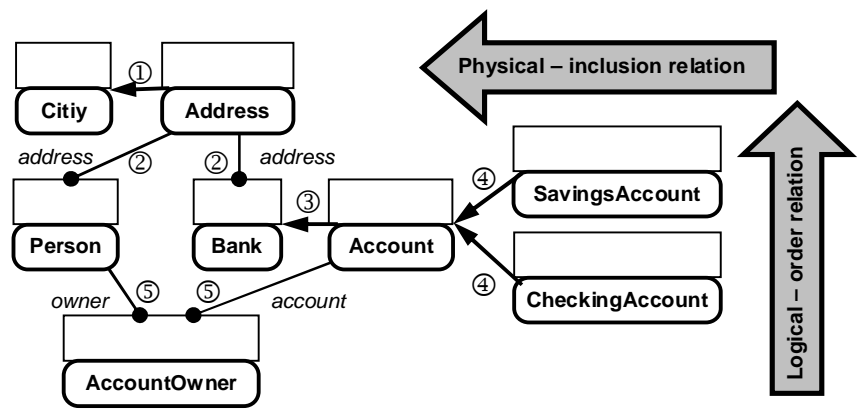

*Figure 5.* Concept-oriented schema.

Concept-oriented schema defines structure of types which can then be used for instantiating collections and instances. Just as concepts, collections exist as members of a partially ordered set. When a new collection is being created we need to *bind* it to the greater collections where its greater elements are stored. This can be done by assigning the corresponding dimensions of the collection when it is being created. For example, a new collection `Persons` can be bound to the greater collection `Addresses`:

```
CREATE TABLE Persons CONCEPT Person, address = Addresses
```

Here we say that elements of the new collection will reference elements of the existing collection `Addresses` in their dimension `address`.

A number of collections bound according to a concept-oriented schema are referred to as a *concept-oriented database schema* (or collection schema). The simplest way to produce a database schema consists in creating one collection for one concept. However, it is also possible to create many collections of one concept or to use one collection by many instances of one of its lesser collections. In data modeling instances exist in collections (in contrast to programming) and for simplicity the terms concept and collection (of this concept) will be used interchangeably. For example, getting instances of some concept will mean getting elements from a collection of this concept.

A set of concept instances existing within a database schema is referred to as a *concept-oriented database*. Just as concepts and collections, instances participate in two structures: any instance has one super-element and a number of greater elements. In queries, the super-element is accessed via keyword 'super' while greater elements are accessed via dimension names.

# 3   Interpretations of Partial Order

In this section we discuss how partial order can be interpreted using conventional data modeling terms and patterns. In the beginning, we will discuss interpretations of one element of order relation represented by one reference stored in a dimension. In the second half, we consider how the whole partially ordered structure of concepts can be interpreted in terms generally accepted in data modeling.

The first interesting observation is that partial order does not provide a built-in mechanism for object characterization via properties, that is, it does not distinguish between objects and their characteristics. The only thing that is formally available is ordering of concepts (schema), collections (database schema) and elements (database). Yet, this ordering can then be interpreted traditionally in terms of object properties or characteristics using the following rule: greater elements are *values* characterizing this element while lesser elements are *objects* characterized by this element. For example (Fig. 6a), if an employee (lesser element) stores a reference to a department (greater element) where this employee works then the employee is an object while the department is a value characterizing this object. One difference from other approaches based on attributes and values is that values in COM are actually reference-object couples where only the reference is a value.

One of the most important interpretations is that greater element is considered *more general* than its lesser elements. Conversely, lesser element is considered *more specific* than its greater elements. As a consequence, if we know that some notion in the problem domain is more specific than another notion then it has to be defined as a lesser concept and positioned under its more general concept. For example, if we know that concept `Product` is more specific than concept `Category` then `Product` is positioned under `Category` (Fig. 6b). Since order relation is represented by means of references, this rule can be reformulated as follows: a referenced element is more general than the referencing elements. It is a highly general and very important rule which interprets a reference as a semantic unit. Additionally, taking into account the previous attribute-value interpretation we can also derive the following rule: a value is always more general than the object it characterizes. For example, if an employee record references its department then this means that the department is a more general element than the employee just because it is referenced.

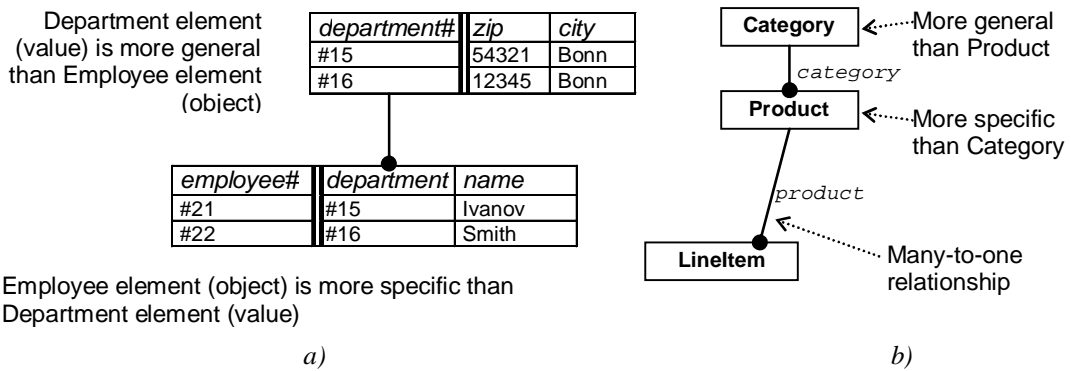

*Figure 6. Interpretations of order relation as attribute-value and specific-general relations.*

A reference representing an element of order relation can also be interpreted as IS-A relationship which means that an element *is* (a more specific case of) its referenced greater element. For example, an employee *is* (a specific case of) the department because references it in its record. This example is somewhat unusual because it is not clear how an employee can be a more specific department but the point here is that IS-A means that a more specific element (employee) can be reduced to a more general element (department) by removing some of its properties. In other words, if we remove all employee properties except for its department then we effectively will reduce its description it to this department and then this employee will be equivalent (IS-A) to the referenced element.

From the point of view of grouping, an element is interpreted as a *combination* of its greater elements, and this combination is then traditionally thought of as an aggregate (Smith et al, 1977), object, record or tuple storing references to greater elements as values in their dimensions. However, an element has also a dual interpretation as a *collection* consisting of its lesser elements (like in the grouping algebra of Li et al, 1996). For example, if we know that a department consists of employees then according to this interpretation employees are lesser elements of the department. In concept-oriented schema this means that concept `Employee` is less than concept

`Department`. Including an employee in a department is equivalent to making it a lesser record than this department record. If a person is a combination of its address *and* project then they are its greater elements positioned over it. Shortly, all constituents of a collection are positioned under it while all constituents of a combination (tuple, record or object) are positioned over it. Thus positioning in a partially ordered set defines the meaning of data. Note that any element and concept can be simultaneously a collection (because it has lesser elements) and a combination (because it has greater elements).

Logically, an element is interpreted as a *conjunction* of its greater elements and a *disjunction* of its lesser elements. In concept-oriented schema this interpretation can be easily represented if all upward edges are thought of as connected by conjunction while all downward connections are connected by disjunction. For example, a department is a disjunction of employees (its lesser elements) while an employee is a conjunction of the department and address (its greater elements).

A simple method for building concept-oriented schema consists in representing all many-to-one relationships between entities by upward arrows in the schema. This means that dimensions in concept-oriented schema are interpreted as many-to-one relationships connecting a source lesser concept (many-part) with the target greater concept (one-part). For example, if *many* products can reference (belong to) *one* category then concept `Product` is a lesser concept of concept `Category` (Fig. 6b). Thus many-part of the relationship is always positioned under the one-part.

Concept-oriented schema can be used to represent the star/snowflake-schema in multidimensional modeling and data warehousing (Berson et al, 1997). These conventional styles of arranging tables assume that there are one or more fact tables referencing many dimension tables. In the case of the snowflake schema, a dimension table can further reference other dimension tables. In COM, the fact table is represented by a lesser concept (frequently it is a bottom concept) while dimension tables are its greater concepts. Thus transformation is reduced to ordering tables as it is shown in Fig. 7. This ordering plays a crucial role in COM because it defines how elements are interpreted and underlies operations with data. In other words, the ordering is not a matter of visual design but rather allows us to bring semantics in the model.

It is important to understand that concept-oriented schema is not a graph where concepts are nodes and dimensions are edges. The true analogy with graphs could be applied if we assume that greater concepts are nodes and lesser concepts are hyper-edges. If a lesser concept has only two dimensions then it is a normal edge connecting two nodes. For example, concept `LineItem` in Fig. 7 is interpreted as an edge between nodes `Order` and `Product`. Interestingly, if we use this interpretation then edges can also be connected by other edges, i.e., such a graph has a layered structure. Primitive concepts are initial nodes of the graph. Lesser concepts of the primitive concepts are edges of the first level. Their lesser concepts are already edges for edges and so on.

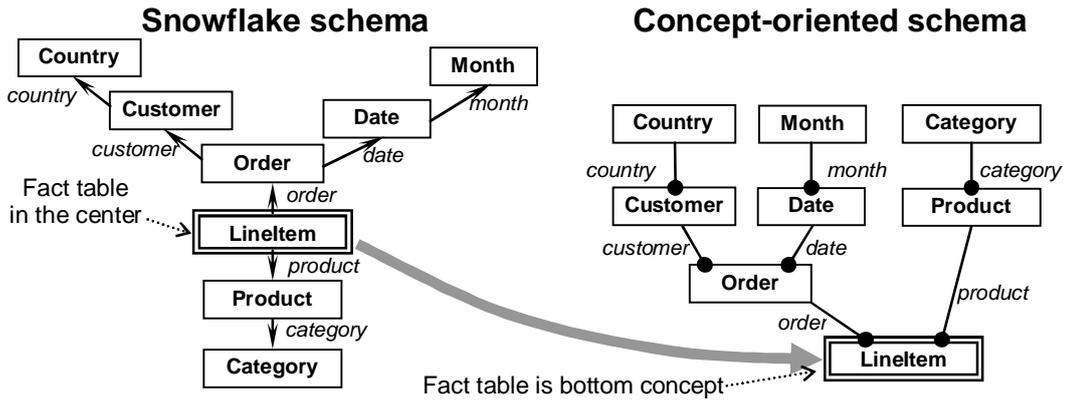

*Figure 7. Snowflake schema and concept-oriented schema.*

One of the most interesting and fruitful interpretations of concept-oriented schema is to treat it as a hierarchical multidimensional space. (Here by hierarchy we mean a hierarchy in the ordered structure rather than inclusion hierarchy.) A classical multidimensional space is defined as the Cartesian product of domains where a *point* in space is represented by a combination of values in these domains, called *coordinates*. Geometrically, one domain is an axis and one point is represented by coordinates along these axes. The main assumption in COM in this context is that a multidimensional space is represented by a lesser concept while its axes (domains) are immediate greater concepts. For example, a two-dimensional space *Z* (Fig. 8a) with two axes *X* and *Y*, $Z = X \times Y$, is described by the concept-oriented schema (Fig. 8b) consisting of three concepts *X*, *Y* and *Z* where *X* and *Y* are two greater concepts of *Z*, $Z <_x X$ and $Z <_y Y$. Concepts *X* and *Y* are treated as domains while dimensions *x* and *y* with these domains are axes or variables. A point in space *Z* is represented by one instance of concept *Z* so that this instance has greater elements representing its coordinates along axes *X* and *Y*. Applying this analogy to the example in Fig. 7, we can say that one order part (instance of concept `LineItem`) is a point in two-dimensional space with the axes represented by concepts `Order` and `Product`.

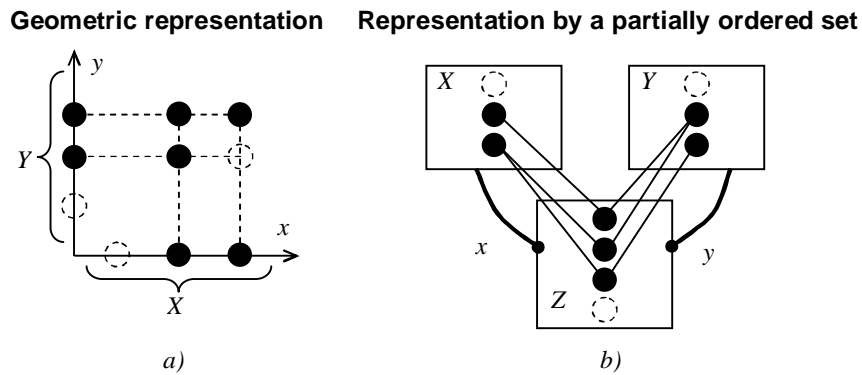

*Figure 8. Representation of a two-dimensional space by a nested partially ordered set.*

In contrast to classical (flat) multidimensional space, COM does not strictly assign the roles of axis with coordinates and space with points to elements of the model. One and the same concept is interpreted as an axis for its lesser concepts and as a multidimensional space with respect to its greater concepts. The consequence is that an existing multidimensional space can be used as an axis for other spaces. For example, concept Z defined in Fig. 8 as a two-dimensional space with two axes X and Y can be used as an independent dimension for its new lesser concept W in Fig. 9a (1). What is new here is that new spaces can use existing multidimensional spaces as domains for their dimensions. On the other hand, if we already have an axis then its coordinates are not necessarily primitive elements but may have their own structure. Such a complex axis is defined as a multidimensional space with its own axes. For example, axis Y in Fig. 9b (2) which was a primitive concept in Fig. 8, is transformed to a two-dimensional space with axes U and V. Thus we can bring a hierarchy into the space structure by extending it upwards.

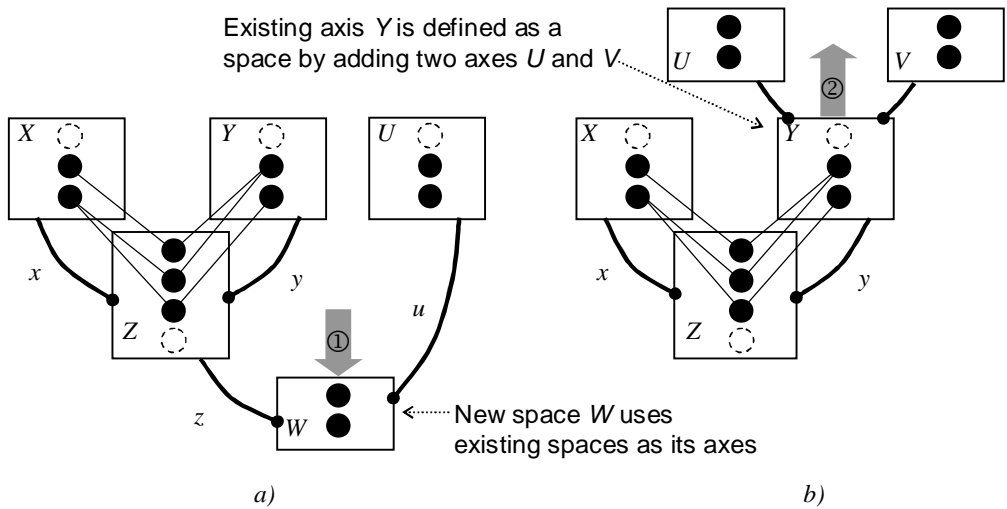

*Figure 9.* *New space can use existing spaces as axes (a) and an existing axis can be defined as a space (b).*

By adding new lesser concepts or greater concepts we can increase the depth of the space and add more levels to its structure. So each point still has some coordinates but these coordinates are now points with their own coordinates. The data modeler thinks of the problem domain as a hierarchical multidimensional coordinate system where existing elements are used to characterize (semantically define) new elements. Another advantage is that such a geometric analogy is very convenient for formally describing data semantics. The idea is that the structure of the model is described by concept-oriented schema while data semantics is a set of points in this space. By adding new points, deleting existing points or changing their coordinates we also change the contents of the database.

One problem with partial order is that it does not allow for self-references (loops) and cycles. For example, formally it is not possible for an employee to reference himself as a manager. There are several reasons for this constraint. Semantically, any new concept is defined via its greater concepts and using self-references or cycles will produce a recursive definition where a term is defined via this same term. Formally, we would like to have a model with a finite number of dimensions where a dimension is defined as a path from bottom to top. In the case of cycles we get an infinite number of dimensions because there will be an infinite number of dimension paths from bottom to top. In the presence of cycles we also are not able to use the inference mechanism (just because elements are defined via themselves). Yet, from practical point of view partial order is too strong constraint which effectively prohibits wide spread patterns. To avoid such problems the requirement of having strict partial order can be weakened by permitting self-references. Such dimensions are then ignored by the mechanisms where partial order is important. Cycles also can be permitted but this issue will be described in some future publication.

## 4 Data Modeling Mechanisms

### 4.1 Logical Navigation via Projection and De-Projection

As we described in Section 2.2, an element is represented by its complex identity consisting of a sequence of segments modeled by concepts and inclusion relation. The main purpose of complex identities consists in providing physical access to the represented element in the hierarchical address space. Therefore, using complex identities and inclusion hierarchy for navigation is referred to as *physical navigation*. By *logical navigation* we mean access to elements using dimensions and partially ordered structure.

Logical navigation is based on two operations of projection and de-projection for manipulating sets of elements.

> **Definition 6 [Projection].** *Projection* is as an operation that is applied to a set of elements and returns a set of their greater elements referenced along the specified dimension.

In COQL, projection is denoted by right arrow '->' and uses three parameters: source collection, dimension name of the source collection and target greater collection from where greater elements are selected:

```
ResultCollection = SourceCollectoin
    -> dimension -> GreaterCollection
```

Here `ResultCollection` is a set of elements from `GreaterCollection` which are referenced at least one time by some element from `SourceCollection` along the specified dimension. In many cases either dimension name or greater collection name can be omitted if they can be unambiguously identified. The result of projection is a set and it includes any element only once even if it is referenced by many elements from the source collection. For example, if we have a set of persons in `Persons` then their addresses can be obtained by projecting this set along dimension `address`:

```
ResultCollection = Persons
    -> address -> Addresses
```

Both source collection and greater collection can be restricted by imposing additional constraints. For example (Fig. 10, left), if only persons with the name starting with 'A' are considered and we are interested in finding only those having zip code '12345' then it is written as follows:

```
ResultCollection = (Persons | name STARTSWITH 'A')
    -> address -> (Addresses | zip == '12345')
```

The result of this query will consist of two addresses #16 and #17.

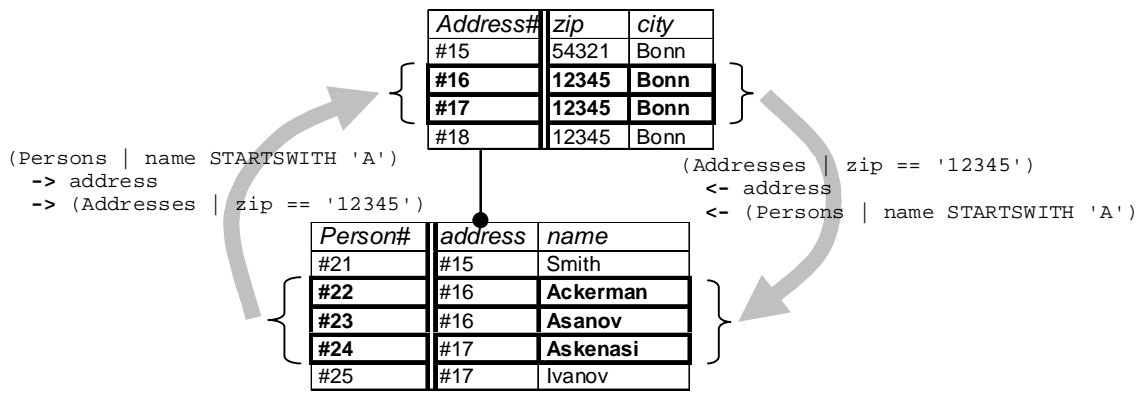

*Figure 10. Projection and de-projection.*

**Definition 7 [De-projection].** De-projection is an operation that is applied to a set of elements and returns a set of their lesser elements that reference the source elements along the specified dimension.

De-projection is denoted by left arrow '<-' and uses three parameters: source collection, inverse dimension (dimension of the lesser collection) and target lesser collection from where lesser elements are selected:

```
ResultCollection = SourceCollectoin
    <- dimension <- LesserCollection
```

Here `ResultCollection` is a set of elements from `LesserCollection` which reference some element from `SourceCollection` along the specified dimension. For example, if we have a set of addresses then the persons living there can be obtained by de-projecting this set along dimension `address` of concept `Person`:

```
ResultCollection = Addresses
    <- address <- Persons
```

If we are interested in only a subset of addresses and persons then they can be restricted as usual:

```
ResultCollection = (Addresses | zip == '12345')
    <- address <- (Persons | name STARTSWITH 'A')
```

This query returns three persons #22, #23 and #24 with the name starting with 'A' and living in area with the specified zip code (Fig. 10, right).

For physical navigation over inclusion hierarchy the same operations of projection and de-projection are adopted. The difference is that the keyword 'super' is used instead of normal dimension names to refer to the physical parent element. In other words, if dimension name points to a logical parent element in the partially ordered set then the 'super' keyword points to the physical parent element in inclusion hierarchy. This type of projection returns a set of all physical parents for the source set of elements:

```
ResultCollection = SourceCollectoin
    -> super -> ParentCollection
```

For example, given a set of accounts we can get a set of their banks:

```
ResultCollection = Accounts
    -> super -> Banks
```

Physical de-projection allows us to get all physical child elements for this set of elements:

```
ResultCollection = SourceCollectoin
    <- super <- ChildCollection
```

For example, we can return all savings accounts with large balance for main accounts with small balance using two de-projection steps:

```
ResultCollection = Banks
    <- super <- (Accounts | balance < 100)
    <- super <- (SavingsAccounts | balance > 100)
```

The real power of projection and de-projection operations comes from their two properties: (i) projection and de-projection operations can be applied sequentially, and (ii) projection and de-projection operations can be used as conditions within other projections and de-projections. A sequence of projection and de-projection operations where each next operation is applied to the result produced by the previous operation is referred to as an *access path*. By *physical access path* we mean an access path which uses only the hierarchical structure induced by inclusion relation. Physical access path involves only the special 'super' dimension in its projections and de-projections and it allows us to navigate horizontally through the database schema. By *logical access path* we mean an access path which uses only the partially ordered structure of collections where dimensions are used for projections and de-projections. Logical access path allows us to navigate vertically by moving up and down through the database schema. The point in access path where it changes its direction between upward and downward (for logical access path) or leftward and rightward (for physical access path) is referred to as a *turn point*.

Suppose (Fig. 11) we have a collection of addresses of persons in Berlin and want to find all related bank accounts for persons older than 20 in banks with address in Bonn. This can be done by de-projecting these addresses for two steps down to `AccountOwners` and then projecting up to `Accounts` with the corresponding intermediate restrictions:

```
ResultCollection = (Addresses | city = 'Berlin')
    <- address <- (Persons | age > 20)                          (1)
    <- owner <- AccountOwners                                   (2)
    -> account -> (Accounts | super.address.city = 'Bonn')(3)
```

It is a logical access path consisting of three segments: two downward segments and one upward segment. Simple constraints are applied to the source (`Addresses`), one intermediate (`Persons`) and the target (`Accounts`) collections.

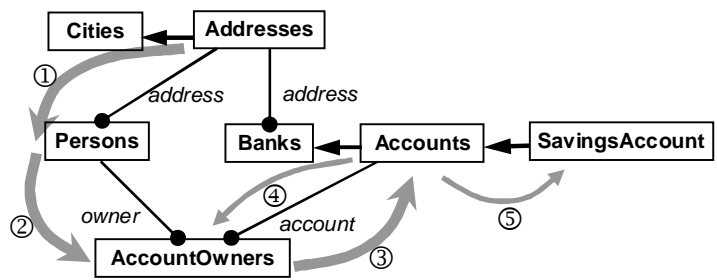

*Figure 11. Access path via projection and de-projection in database schema.*

Let us now make this example more complicated by adding de-projection with aggregation to intermediate constraints. For example, we might be interested in finding only accounts with at least two owners and at least 100 EUR on all their savings accounts. In this case the last projection of the above query is modified as follows:

```
-> account -> (Accounts |                                       (3)
    super.address.city = 'Bonn' AND
    this <- account <- AccountOwners > 2 AND                    (4)
    SUM(this <- super <- SavingsAccounts.balance) > 100         (5)
    )
```

Here de-projection `this <- account <- AccountOwners` returns all account owners for this account and comparison with 2 is a shortcut for the collection size (`COUNT` aggregation function). The last condition selects all savings sub-accounts `this <- account <- AccountOwners`. For these savings accounts we are interested only in one numeric field `balance`, which is used in the `SUM` aggregation function.

This example shows how rather complex queries can be written in a very concise and natural form. Importantly, access path does not depend on the identification mechanism, i.e., we can change the format of references while all the queries will remain unchanged. This is possible because access path queries do not involve any information about implementation of connections but include only dimension names. For comparison, SQL queries mix these two concerns and a query normally needs to specify not only what we want to get but also numerous details about how it can be obtained using joins.

## 4.2 Multidimensional Analysis

In the previous section we described an approach to data access based on access path. However, in many applications such as multidimensional analysis and OLAP we need to produce a *new* collection as a product of existing ones rather than to retrieve existing elements. In this case many source collections are used to build one result collection as their Cartesian product, i.e., one result element is a combination of some source elements. In COQL this operation will be written using round brackets prefixed with the keyword CUBE and the source collections separated by comma. For example, we could build a new collection as a product of all cities and all banks:

```
ResultCube = CUBE ( Cities city, Banks bank )         // Product
```

Each element of a new multidimensional collection is treated as a cell of a multidimensional cube. If we need to restrict elements in the result cube then it is done as usual by specifying a condition all its elements have to satisfy. The condition is either separated from the source collections by bar symbol or is prefixed with the keyword WHERE. For example, the following query will use only banks with the name starting with 'A' for building the cube:

```
ResultCube = CUBE ( Cities city, Banks bank )
    WHERE ( bank.name STARTSWITH 'A' )
```

The main idea of multidimensional analysis is that for all elements of the cube, we have to compute some value which is referred to as a *measure*. Actually, there can be many such computed values assigned to cells of the cube and they need not be numeric. The measure can be added as an additional dimension of the result collection but for that purpose we need to have a possibility to change the structure of the result collection. In COQL, this is done by means of the keyword RETURN which is analogous to SELECT in SQL. The RETURN keyword specifies the values for each cell that need to be included in the result collection. By default (if this keyword is omitted) the values of the source collections are returned:

```
ResultCube = CUBE ( Cities city, Banks bank )
    RETURN ( city, bank )                  // Result dimensions
```

There are two differences between how product is implemented in COQL and SQL:

- [Arity of the result] The number of columns in the result (arity) produced by SQL product is equal to the sum of columns in the input tables specified in FROM clause. In COM, arity of the result is equal to the number of input collections specified in CUBE.

- [Cardinality of the result] SQL product is projected to the selected attributes listed in SELECT clause and hence the cardinality of the result does depend on the selected attributes. In COQL the number of elements in the product (cardinality) does not depend on the returned variables listed in RETURN.

If we need to include some measure then it has to be computed and then returned along with the source elements. Simple measure can be computed directly within the RETURN clause:

```
ResultCube = CUBE ( Cities city, Banks bank )
    RETURN ( city, bank, measure = bank.accs / city.pop )
```

Here the measure is computed as the number of accounts in this bank for each citizen in this city. The result collection is a three-dimensional space where the first two dimensions are independent while the third one is a measure.

In order to compute more complex measures it is useful to have intermediate reusable local variables which can be computed within a special query block marked by the keyword BODY. For example, using the BODY block, the above query can be equivalently rewritten as follows:

```
ResultCube = CUBE ( Cities city, Banks bank )
    BODY ( measure = bank.accs / city.pop )      // Query body
    RETURN ( city, bank, measure )
```

The BODY block can be thought of as being computed for each iteration over all cities and banks restricted by the constraints imposed in the WHERE clause. Yet, the order of computations is not determined and it is a normal set operation because we know only that the source instance variables are assigned within this block but do not know the previous cell and the next cell during the computation.

Suppose that we want to build a diagram with cities and banks as horizontal axes. As a measure, this diagram has to draw the total account balance of all persons from this city owning an account in this bank. Note that a person may have accounts in many banks and one account can be owned by many persons living in different cities. Data for this diagram can be produced using the following query (Fig. 12):

```
CUBE ( Cities city, Banks bank )                         // Dimensions
BODY (
    CityAccounts =
        city <- super <- Addresses                       (1)
        <- address <- Persons                            (2)
        <- owner <- AccountOwners                        (3)
        -> account -> ( Accounts | parent.bank == bank ) (4)
    measure = SUM( CityAccounts.balance )                (5)
)
RETURN ( city, bank, measure )
```

Here we start from all addresses in the `city` (1) and find all its account owners by de-projecting to `Persons` (2) and then to `AccountOwners` (3). Then these account owners are projected to `Accounts` (4) but we select only those in the current bank. Finally, we need to aggregate data in this intermediate collection by summing up their balance and storing it in variable `measure` (5) which is returned as the third dimension.

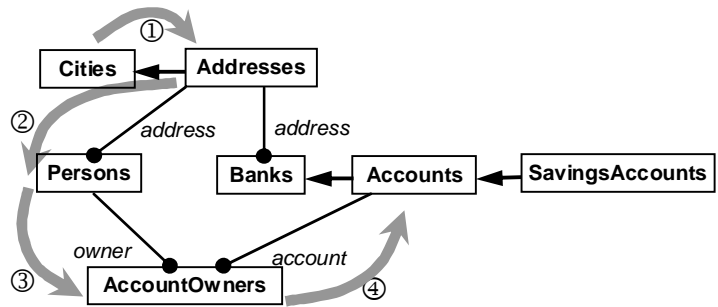

*Figure 12. Finding a group of elements for one cell in a cube.*

The type of analysis where elements are viewed simultaneously along several dimensions each having many levels of details is studied in online analytical processing (OLAP). In the concept-oriented model, it can be performed using a procedure consisting of the following steps:

1. Choose one *fact concept* consisting of elements that will be grouped along several dimensions and impose constraints on its elements
2. Choose several *dimension paths* each starting from the fact concept and then proceeding to its greater concepts
3. Choose one *level concept* along each dimension path and impose constraints on them
4. Build a *multidimensional cube* as the Cartesian product of all the level concepts
5. Group elements of the fact concept over elements of the multidimensional cube
6. Choose a measure of the fact concept and compute its aggregated value for each group

The result of such an analysis will be one aggregated measure property associated with one cell of the multidimensional cube.

Let us assume that we need to analyze how the company sales are distributed in the space of customers and products. Each individual sale is stored in the `LineItems` collection (Fig. 13). Elements of this collection will be grouped over two dimensions and some property of each element will then be aggregated. Hence we choose `LineItems` as a fact collection in our analysis.

The first dimension path describes customers:

```
LineItems -> order -> customer -> country    // (1) in Fig. 13
```

Any fact (one sale) is characterized by one order it belongs to and one customer or a country, depending on the level of detail we choose on the next step. The second dimension path describes products:

```
LineItems -> product -> category             // (2) in Fig. 13
```

Now any fact (one sale) is characterized by two values: one from the customer dimension path and the second from the product dimension path.

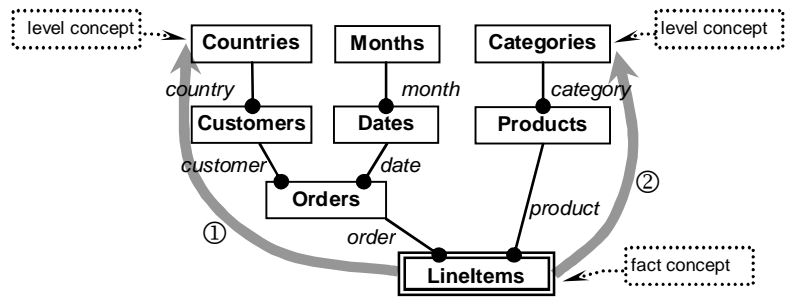

*Figure 13. An example of multidimensional analysis.*

Each dimension path chosen on the previous step consists of several collections which differ in their level of detail. For analysis, we need to choose one level of detail for each dimension path. For example, we might start from the lowest level of detail by choosing `Countries` as a characteristic of customers and `Category` as a characteristic of products. If later we need to see more details then we can drill down to a lesser collection along the same dimension path. After choosing these level collections, each fact element belongs simultaneously to a pair of level elements. In other words, one sale belongs to one country *and* one category as the groups.

Now let us consider how facts are grouped over elements of multidimensional space. First of all we need to produce a multidimensional space from the chosen level collections:

```
CUBE ( Countries co, Categories ca )
```

The result of this query is a collection consisting of all combinations of countries and categories. If we want to consider only countries with at least one customer then the size of the cube is restricted by imposing the corresponding constraints:

```
CUBE ( Countries co, Categories ca )
WHERE ( co <- country <- Customers > 0 )
```

Here we de-project this country and compare the size of the group of obtained customers with zero. Here again we used a shortcut which in full form is equivalently written using the aggregation function:

```
WHERE ( COUNT(co <- country <- Customers) > 0 )
```

Now it is necessary to find fact elements associated with each cell of the two-dimensional cube. This can be done by finding a subset of order elements projected to both the current country and category:

```
cellGroup = CUBE LineItems op
    WHERE op -> order -> customer -> country == co AND
          op -> product -> category == ca AND
          op.date == 2007
```

Note that we also added an additional constraint by selecting only facts from 2007. Using intersection of two de-projections the same can be written as follows:

```
cellGroup =
    co <- country <- customer <- order <- LineItems AND
    ca <- category <- product <- LineItems AND
    (Dates | year == 2007) <- LineItems
```

Here three de-projections start from instances on different levels and end in the same fact collection `LineItems`. The fact elements associated with each cell of the cube are found as an intersection of all one-dimensional de-projections. Note that AND operator is overloaded: it is interpreted either as a logical connective or as a set intersection.

Once we have found a group of fact elements for each cell of the two-dimensional cube it is now necessary to define the measure which is some aggregated property of this group. For example, we might sum up the price paid for the orders within one group:

```
totalPrice = SUM ( cellGroup.price )
```

It is possible to select the second measure as the number of orders in the group of order parts:

```
orderCount = COUNT ( cellGroup -> order )
```

These measures are then included in the query output via `RETURN` clause:

```
RETURN co.code, ca.id, totalPrice, orderCount
```

The whole query is written as follows:

```
CUBE ( Countries co, Categories ca )
WHERE ( co <- country <- Customers > 0 )
BODY (
    cellGroup =
      co <- country <- customer <- order <- LineItems AND
      ca <- category <- product <- LineItems AND
      (Dates | year == 2007) <- LineItems
    totalPrice = SUM ( cellGroup.price )
    orderCount = COUNT ( cellGroup -> order )
    )
RETURN co.code, ca.id, totalPrice, orderCount
```

In the case we need more detailed analysis it is possible to choose other level collections along the dimension paths. If we move down in the schema and choose a lesser collection then this operation is equivalent to drill down. If we move up and choose a greater collection along this dimension path then this operation is equivalent to roll up.

## 4.3 Inference

One application of projection and de-projection operations consists in automatically producing a result set with related elements without the necessity to specify an exact access path with projections and de-projections. For example, let us consider an example shown in Fig. 14 where two collections `Orders` and `Products` have one common lesser collection `LineItems` which is treated as a dependency between its greater collections. The presence of dependency means that if we select some orders then this entails selection of related products and vice versa if some products are selected then we can infer related orders. Of course, the related data elements can always be obtained by providing a concrete query with precise rules how the source constraints need to be propagated to the target set. However, the presence of ordering allows us to propagate constraints automatically using the following two-step procedure:

1. [de-projection phase] source constraints are propagated down to lesser collections using de-projection, and then

2. [projection] the constrained lesser collections are projected up to the result collection(s)

The first step is denoted by left arrow (de-projection) with star suffix '<-*' where the star means any downward path to some lesser collection. The second step is denoted by right arrow (projection) with star prefix '*->' where the star means any upward path from the obtained result to the specified greater collection. Inference operator is denoted as '<-*->' which is interpreted as 'first de-project and then project'.

Assume that the question is what orders are related to beer and chips, that is, we need to find all orders where either beer or chips are product items. Using this operator this query is written as follows:

```
RelatedOrders = ( Products | name IN {'beer', 'chips'} )
    <-*-> Orders
```

Operator '<-*->' denotes inference (constraint propagation) from the collection `Products` to the second collection `Orders` using their common lesser collections. Note that this query provides no indication how the first (restricted) collection `Products` is connected with the second collection `Orders`. What is interesting, this information is not needed because the constraints can be propagated automatically. The selected two products, beer and chips, will be de-projected to the lesser collection `LineItems` which will contain only three elements. Then these three elements are projected to the collection `Orders` and two rows #23 and #24 will be returned as a result of this query.

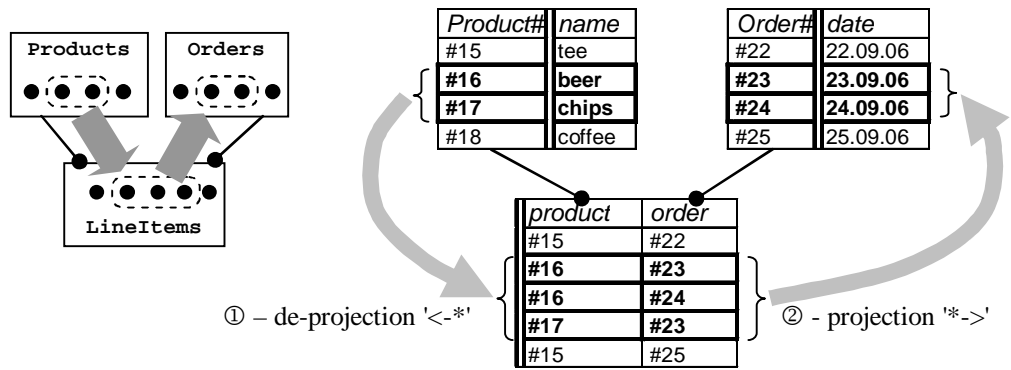

*Figure 14. Inference via constraint propagation.*

In the case of many constraints in different parts of the database schema they are propagated down to the most specific collection. On the second step, their intersection is aggregated up to the target set. For example, a model in Fig. 15 consists of three already described collections `Orders`, `Products` and `LineItems`. However, `Orders` and `Products` have their own greater collections. In particular, each order is characterized by a customer (who made this order) and a date (when this order was made). Each customer belongs to some country from set `Countries` and each product has a category from collection `Categories`. Let us now assume that we want to get all countries related to some product category (say, `'cars'`) and during some period of time (say, in `'June'`). In other words, we want to learn in what countries cars were sold in June. This can be done by the following query where we explicitly specify the constraint propagation path (dimension names are omitted for simplicity):

```
RelatedCountries = (
    ( Categories | name == 'cars' )
        <- Products <- LineItems AND      // Path (1) Fig. 15
    ( Months | name == 'June' )
        <- Dates <- Orders <- LineItems   // Path (2) Fig. 15
    )
    -> Orders -> Customers -> Country     // Path (3) Fig. 15
```

However, this query is not only long but also depends on the concept-oriented schema. Using inference operator it is possible to impose our constraints and indicate what kind of result we want to get — all the rest will be done automatically:

```
RelatedCountries =
    (Categories | name == 'cars') AND
    (Months | name == 'June')
        <-*-> Countries
```

The database engine will effectively remove all non-car items from the database and produce some subset of all available order parts. The second constraint consists in selecting only items characterized by June as their date. When this constraint is propagated down, all non-June items from its subsets are effectively removed. In particular, all non-June dates, all non-June orders and all non-June order parts do not satisfy this constraint. After that set `LineItems` will contain only elements characterized by cars as its product category and by June as its date. The last step in this procedure consists in aggregating the selected order parts up to the target collection `Countries`.

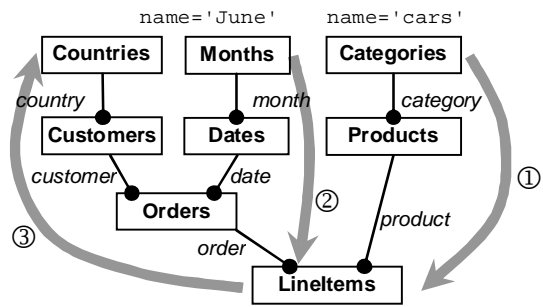

*Figure 15.* *Automatic propagation of two source constraints.*

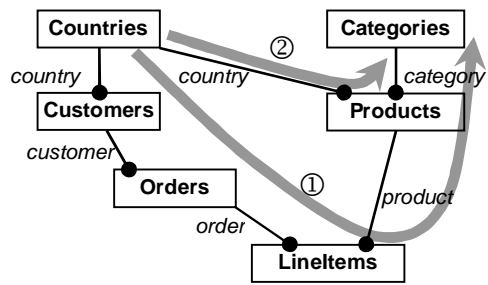

*Figure 16.* *Alternative inference paths.*

The described procedure assumes that the source constraints are propagated down along all inverse dimensions. However, if the model has many possible access paths then it is necessary to specify which one to choose in the query. For example (Fig. 16), collection `Countries` can be a greater collection for `Customers` and `Products`. If we want to get related product categories for a selected country then there exist two options for constraint propagation from `Countries` to `Categories`. The first path (1) goes through collection `LineItems` and this inference strategy will result in all product categories *ordered* by customers from the specified country. The second path (2) goes through collection `Products` and it will return all categories for products *made* in the specified country. To avoid default propagation along both paths we can provide an intermediate collection as a parameter of the operator between star symbols which as usual denote an arbitrary dimension path. The first query below will follow path (1) in Fig. 16 while the second query will follow path (2):

```
RelatedCategories =
    ( Countries | name == 'Germany')
        <-*(LineItems)*-> Categories // (1) Via LineItems
RelatedCategories =
    ( Countries | name == 'Germany' )
        <-*(Products)*-> Categories         // (2) Via Products
```

Another difficulty which can arise during inference is the absence of a common lesser collection for the source and target greater collections. In this case constraints propagated down along all inverse dimensions cannot be aggregated because they do not reach the target collection. For example, a model in Fig. 17 assumes that coaches (collection `Coaches`) train teams (`Teams`) while one team consists of a number of players (`Players`). Collections `Trains` and `Plays` store pairs of coach-team and player-team, respectively. Thus a player may play for many teams and a coach may train many teams. Now let us try to ask the following question: find players related to a selected coach. For this model this question means that we want to get all players who have ever played for any team trained by this coach. As usual, this problem can be solved by specifying manually an exact access path:

```
RelatedPlayers = ( Coaches | name == 'Klinsmann' )
    <- Trains -> Teams
    <- Plays -> Players
```

(Here we again specify access path using only collection names without dimensions.)

Here the operator '<-*->' does not work because there is no path from `Coaches` to `Players` (they do not have a common lesser collection). Indeed, if we select the coach and propagate this constraint down according to the first step of our procedure then we get collection `Trains`. Here this procedure stops because there is no path leading to the target collection `Players`. We can project `Trains` to its greater collection `Teams` but not to `Players` because players are not directly connected with coaches.

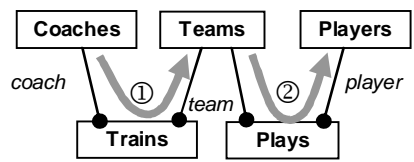

*Figure 17.* *Zigzag propagation path.*

One solution could be using two operators consecutively:

```
RelatedPlayers = ( Coaches | name == 'Klinsmann' )
    <-*-> Teams <-*-> Players
```

The second solution is based on the observation that the first collection (`Coaches`) and the last collection (`Players`) have a common bottom collection which can be introduced formally (Fig. 18). In this case we can perform standard inference but the result will be wrong because bottom collection stores no data. Therefore, the following query will return *all* players independent of the selection of coaches:

```
RelatedPlayers = ( Coaches | name == 'Klinsmann' )
    <-*-> Players
```

To solve this problem we can use background knowledge which is absent in the database but is assumed in our query. Indeed, if we want to get all players trained by some coach then we implicitly assume that the coach trains a team of this player. In other words, the team trained by the coach must be the same team where the player plays which is expressed as follows:

```
( Trains.team == Plays.team )
```

If this assumption could be explicitly formulated in the query then the source constraints would be correctly propagated from coaches to players via formally added bottom collection (Fig. 18). This problem can be solved by imposing the implicit constraints on bottom collection:

```
( Bottom | trains.team == plays.team )
```

This restriction means that a coach is related to a player only if he trains the same team where this player is a member. As a result, bottom collection (formally) contains only elements which satisfy this additional condition rather than all possible combinations of its two greater collections. Hence it now contains the necessary dependency that will be taken into account during inference:

```
RelatedPlayers = ( Coaches | name == 'Klinsmann' )
    <-*(Bottom | Trains.team == Plays.team)*-> Players
```

In other situations this additional constraint could express some other background knowledge about the problem domain. For example, a player might be treated as related to a coach if he was playing in the team for more than a year or more than some number of games.

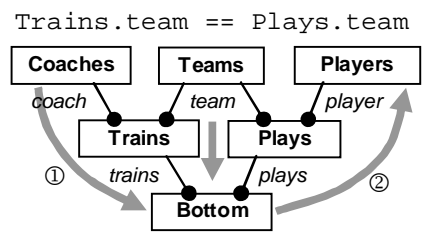

*Figure 18. Inference using background knowledge*

## 5 Related Work and Discussion

### 5.1 Identities and Identity Modeling

One of the main distinguishing features of COM is that it makes identity and entity modeling symmetric by providing a mechanism for describing domain-specific identities (references). Identification mechanisms can be divided into two big groups (Khoshafian et al, 1986; Eliassen et al, 1991): (i) strong identities, and (ii) weak identities. The main distinction between them is that strong identities are separable from the represented entity while weak identities (identifier keys) is a method where some part of entity (a subset of its properties) is used to represent it. For example, cells in memory and objects in OOP are identified by strong identities. In contrast, primary key in the relational model is a weak identity. COM relies exclusively on strong identities but in contrast to object-oriented approach, they can be explicitly modeled precisely as it is done for entities, i.e., both identities and entities in COM may have domain-specific structure and behavior. This means that COM allows for creating an arbitrary domain-specific address space which works as conventional memory but is an integral part of the model.

Primary keys are not equivalent to identity classes because they are still defined in terms of normal columns. Formally, primary key is an integrity constraint but its main purpose is row identification in terms of row (entity) properties. In contrast, identity class in COM defines identities which exist separately from the represented entity and, strictly speaking, not stored along with the entity properties. Thus primary keys can be characterized as a method of distinguishing rows using these same rows rather than strong identities. In particular, primary key columns may well be changed as simply as any other column of the entity while changing a COM identity is a completely different procedure which can be compared with changing an object reference or a memory cell address.

An important property of COM identities is that even though they are explicitly modeled and may have an arbitrary domain-specific structure, they are used transparently without exposure of this structure. This property is true for object-oriented models and is false for the relational model. As a result, queries in COM are much simpler than in SQL and do not depend on the structure of identities while SQL queries need to be updated when identification scheme changes.

Just as in object-oriented models, COM uses strong identities. The main difference is that are *user-defined* (custom) identities rather than being provided by the underlying environment as some kind of platform-specific identifier (also called primitive, atomic or system identities). Another unique feature is that identities are modeled only as part of one element along with its entity while in object-oriented and other approaches identities can be modeled (using classes or other conventional means) in isolation from the represented entities. For example, we could define a class `BankBIC` (identity) and then a separate class `Bank` (entity). However they will be still normal classes defined separately. The underlying system is unaware of their specific roles and hence cannot help in managing this layer of functionality.

Importantly, COM collections are different from relational tables. The main reason is that collections involve *two* constituents in their definition which play dual roles and cannot be reduced to each other. In relational terms, a collection is a table with columns defined by only the entity class while the identity class is used for describing rows. In other words, entity class is used to define horizontal coordinates of the table while identity class is used to define vertical coordinates of the table. In this sense, concepts provide a symmetric tool for defining the structure of two-dimensional tables while relational tables provide only means for defining their horizontal structure (columns). In relational model there are no means for defining domain-specific row identifiers although normally DBMS provide some kind of primitive row identifier. Formally, the relational model does not even recognize the need in such row identifiers by making a fundamental claim that everything should be done only by means of attributes. Row identifiers are assumed to be part of the physical level and the relational model was designed as a means of abstracting from these details. In this sense, COM not only reverses the situation and brings row identifiers back into the model from underground by legalizing their status (like it was in the hierarchical model): COM goes further and makes the opposite assertion that row identifiers must be an integral part of any good data model at any level of abstraction – a model without strong identities is incomplete, independent of the level of abstraction. COM identities can also be thought of as surrogates with an arbitrary user-defined format which is modeled along with the format of entities.

COM collections can be viewed as a memory model where both cells and addresses can be defined by the modeler using one construct. Note that both cells in memory and entities within a collection do not store their address and identity as their contents. Instead, addresses are managed by a separate mechanism. In particular, updating an address has no consequences because addresses are not stored in memory but rather are implemented by the hardware. In COM it is theoretically possible to change identities but it can be rather complicated procedure which has nothing to do with updating dimensions of entities. The mechanism of concepts can be viewed as a method to define a memory or container with a domain-specific address space structure and domain-specific structure of its elements. The explicit distinction between memory locations and their content was used in the logical data model (LDM) (Kuper et al, 1984; Kuper et al, 1993) by means of object values (r-values) constituting data space and object names (l-values) constituting address space. The distinguishing feature of COM is the use of concepts for modeling these two constituents as two parts of one whole.

One issue within the relational model is that it needs data types for defining the (immutable) values and this aspect is considered an orthogonal direction to the relational algebra. COM and concepts provide an elegant solution to this problem of domain and type modeling by covering simultaneously two orthogonal directions. Such domain and data type modeling is based on using identity class for describing values. Using concepts with only identity class we can define domains with values having an arbitrary application specific format. Note that domains and their values will not be persistent by themselves because they are defined via identity class. But of course they can be stored in entity fields just like normal references and the only difference is that such values do not represent any entity. The distinguishing feature of COM is that identity and entity classes exist only within one concept and hence they are not independent classes (otherwise we could use simply two separate classes) but rather one integral construct which behaves differently in different situations.

## 5.2 Graph-Based and Multidimensional Modeling

The use of inclusion hierarchy in COM is very similar to the hierarchical data model (Tsichritzis et al, 1976) where elements also exist within a hierarchy. The difference between them is that the purpose of the COM inclusion hierarchy consists in describing how elements are represented within a hierarchical address space. In other words, inclusion relation in COM is a means for describing a structured container with a hierarchical address space. Note that this address space can be (and should be) explicitly modeled by providing domain-specific identities instead of platform-specific handles.

Logical navigation in COM makes it similar to many graph-based models like the network model (Taylor et al, 1976), the logical data model (Kuper et al, 1984; Kuper et al, 1993), the functional data model (Shipman, 1981; Gray et al, 1999; Gray et al, 2004), object-oriented models (Dittrich, 1986) and many others (e.g., Kerschberg et al, 1976; Angles et al, 2008). In graph-based models schema and instances are modeled as a graph while data is manipulated via the corresponding navigational operations. The main difference of COM is that it uses partially ordered sets rather than graphs (so COM is not a graph-based model at all). Logical navigation in COM is based on projection and de-projection operations in a partially ordered set while navigation in graph-based models is a kind of 'follow a link' approach. If links in graph-based models are simply navigational tools where the fact of referencing has no semantic interpretation then a reference in COM is an elementary semantic unit. Therefore, navigation paths in COM have a zigzag form where we move up and down between more general and more specific elements, between objects and values, between collections and combinations and so on (see Section 3). Note that using directed acyclic graphs does not change this situation because it is still a graph rather than a partially ordered set.

Data models can be distinguished on their support of nested sets and set-valued attributes. It is very important mechanism because most problem domains have nested structure while sets of values can be used as characteristics of other elements. The simplest existing solution consists in directly introducing multi-valued attributes into the model so that a table column or class field is marked by a keyword like 'multivalued' in SDM (Hammer et al, 1978, 1981). For example, if a department has many employees then the set of employees can be declared as a field using the keyword 'SET OF':

```
CLASS Department
    CHAR[256] name
    SET OF Employee employees        // Multi-valued attribute
```

This natural solution can be convenient in simple models but it entails numerous problems in complex models with complex relationships and what is worse is that it is extremely difficult to formalize. For example, class `Employee` in the above example could also have set-valued attributes with values in other classes including class `Department`. One direction for formalizing data manipulations in this case was developed within nested relation model (Jaeschke et al, 1982; Abiteboul et al, 1986; Schek et al, 1986; Roth et al, 1988; Chen et al, 1991) which is an extension of the relational model.

Although having set-valued attributes seems like a very attractive and useful feature, COM does not adopt it, that is, strictly speaking COM does not have set-valued attributes. Instead, COM proposes to use dimensions with inverse direction for that purpose. In other words, one and the same dimension looks as a single-valued attribute from the referencing class and (implicitly) as a multi-valued attributed from the referenced class. Thus inverting a dimension turns it into a multi-valued attribute. For example, if a department is known to have many employees then we do not need to declare any field in this class at all. Instead, we declare this field in the `Employee` class and this field in inverted form can then be used from the `Department` class:

```
CONCEPT Department
  ENTITY
    CHAR[256] name
    // <- dept <- Employee     Implicit multi-valued attribute
CONCEPT Employee
  ENTITY
    Department dept                  // Single-valued attribute
```

Now if we need to get all employees of one department then we simply use de-projection:

```
(Department | name == 'SALES') <- dept <- Employee
```

It is rather elegant solution which does not add complexity to the model and reuses its ordered structure. Of course, it is always possible to explicitly define a multi-valued attribute as a method of the concept which can be then used in other queries:

```
CONCEPT Department
  ENTITY
    CHAR[256] name
    getEmployees {         // Explicit multi-valued attribute
        RETURN this <- dept <- Employee
        }
```

There exist numerous approaches to modeling multidimensional data which are mainly driven by the demand from analytical applications and data warehouses (Berson et al, 1997) where the notion of dimension and data cube (multidimensional space) plays a primary role (Li et al, 1996; Agrawal et al, 1997; Gyssens et al, 1997; Nguyen et al, 2000; Torlone, 2003; Malinowski et al, 2006). The main difference of COM is that concepts do not have predefined roles of cube, dimension or measure. These roles are assigned later for each concrete analysis so that a concept can be a cube for one task and it can be a dimension level (category) for another task. However, in contrast to traditional OLTP models, COM provides all the necessary facilities for performing various analysis tasks and in this sense it can be viewed as a combined OLTP-OLAP model.

## 5.3 Semantic Models

One of the main characteristics of any semantic data model is its ability to represent complex relationships among elements (Hull et al, 1987; Peckham et al, 1988; Potter et al, 1988) which can then be used for performing complex data operations. The role of semantics in databases is analogous to that in web: in both semantic data models and the Semantic Web (Berners-Lee et al, 2001, May) the idea is to make it possible for the managing system to understand its data and satisfy the requests of users and machines.

The use of references in COM as an elementary semantic construct makes it similar to a class of binary semantic models (Abrial, 1974) which rely on one binary relation for representing meaning. The difference of COM is that references are used to define partial order. Another difference from binary models is that COM references have a hierarchical structure because they are intended for representing elements in a hierarchical address space.

Semantic data models from the second class rely on a rich set of semantic relationships for describing data meaning (Hammer et al, 1978; Hammer et al, 1981; Kent, 1979; Abiteboul et al, 1987; Jagannathan et al, 1988). Most such models are graph-based with links having meaningful roles which are interpreted as relationships arising frequently in typical database applications. Thus data semantics is stored in these relationships and possibly also in inference rules which describe how these relationships have to be used to process data (to query existing or infer new data). The most important abstraction mechanisms of conventional semantic models such as aggregation, classification and generalization can be easily implemented in COM but they lose their rich interpretation because the intention of COM is to decrease the number of elementary semantic constructs. What is most important is the presence of two hierarchies: one induced by inclusion relation and one induced by partial ordering.

## 5.4 Formal Methods

The role of partial order relation in data modeling was studied in (Zaniolo, 1984; Buneman et al, 1991) where partial order is a consequence of having incomplete information in data. The idea is very simple and natural: if a record has no value in some field then it is more general than the same record with any concrete value in this field and, vice versa, a record with an additional attribute is more specific than the original record where it is absent. However, this research was restricted by the frames of the relational model and did not produce new foundations for data modeling. In contrast, COM assumes that data is intrinsically partially ordered as one of its fundamental principles and then this formal order-theoretic setting is used to developing a data model by adding the necessary operations, features and mechanisms.

Another extension of the relational model of data is the universal relation model (URM) (Kent, 1981; Fagin et al, 1982; Maier et al, 1984). This direction was aimed at introducing a kind of canonical representation in the form of a universal relation so that all relations are viewed as its projections. The model would then be viewed as one whole rather than a flat set of unconnected relations with many advantages like global operations with data, global consistency, inference etc. However, the assumption of universal relation was shown to be incompatible with many aspects of the relational model and it did not result in a new foundation for data modeling. Interestingly, COM also introduces a kind of universal relation but makes it on order-theoretic basis and in this sense it achieves the general goals of URM using different means. More specifically, bottom collection in COM can be viewed as an analogue of the universal relation where all dimension paths leading to primitive concepts are interpreted as primitive attributes. The difference is that the universal relation is defined over a set of attributes while the structure of the concept-oriented model has many levels within a partially ordered set. Using inference procedure described in Section 4.3 we can successfully solve problems very similar to those studied in URM where it is necessary to automatically find related data items.

From the point of view of using partially ordered sets for representing data semantics, COM is very close to formal concept analysis (FCA) (Ganter et al, 1999; Wille, 2006) which is a lattice-theoretic method of data analysis. In FCA, initial data is represented as a formal context by means of a set of objects, attributes and incidence relation between them. The task consists in deriving formal concepts each of them represented by a subset of attributes and objects satisfying certain natural conditions. In this sense COM and FCA have different goals and different mechanisms. What is similar is the powerful mathematical formalism of lattices with its rich semantic interpretations that are highly relevant in data and knowledge processing.

Semantic relationships encoded in a data model are intended to be used for automatic data management such as query processing, consistency support, inference and other tasks. One of the most important tasks (and essentially the criterion for determining if a model is really semantic) is inference. Typical systems with inference like deductive databases, concept graphs, ontologies (Gruber, 1993) or the Semantic Web (Berners-Lee et al, 2001, May) assume the existence of inference rules which encode the necessary logic of formal reasoning. In other words, inference rules allow us to express how constraints have to be propagated through the model so that given some input data we can produce the output. In contrast to such models based on formal logic, COM does not need inference rules for carrying out inference. All information that is necessary for inference is encoded directly in partial order relation. In other words, it is the ordered structure that provides natural directions for constraint propagation.

# 6  Conclusion

In this paper we have described a new approach to data modeling and the corresponding query language. The model and query language are based on three principles of duality, inclusion and order which are summarized as follows:

- Duality principle postulates that an element is an identity-entity couple which is modeled by means of concepts. Identities are used to manifest the fact of existence of the represented entity. This principle brings strong domain-specific identities into the focus of data modeling by giving them equal rights with entities. Another its role is that it makes domain and type modeling integral parts of data modeling.

- Inclusion principle postulates that any element exists in space which is also a normal element. To model this space membership we define inclusion relation on concepts. Parent elements play a role of container where child elements exist. This principle turns any element into a set so that the notion of set is supported by the model at the core level. It also allows us to describe domain-specific hierarchical address spaces as integral part of the data modeling process. Inclusion relation generalizes inheritance and this makes this model much closer to object-oriented methods.

- Order principle postulates that a set of elements is partially ordered and this partial order represents their semantics. Partial order among concepts is represented by their dimension types. This structure is then used to describe various mechanisms and patterns existing in data modeling. In particular, two main operations of projection and de-projection are defined in terms of partial order. An important feature of this approach is that is set-based and join-free.

To demonstrate how COM works under "field conditions" we applied it to the following three highly general tasks:

- Logical navigation in COM has been shown to be rather simple, natural and flexible for describing highly complicated queries. The main distinguishing feature of this approach is that it is based on projection and de-projection operations which change the level of detail of the set of data elements.

- Multidimensional analysis is based on product operation which is applied to the chosen level collections within their dimension paths. The advantage is that COM supports both transactional and analytical views of data. The difference from the existing models is that the roles of cube, dimension, levels and measures are assigned for each particular analysis rather than are predefined in its structure.

- Inference is a procedure for reasoning about data by imposing source constraints and then deriving new constraints in other parts of the model. This method is based on a two-step procedure where imposed constraints are propagated down towards more specific concepts using de-projection and then the result is propagated up towards the target more general concept.

In future, we are going to further develop this model in the following major directions:

- Finalizing the formal setting for COM which is called nested partially ordered set. If in this paper we describe this model using a query language then this new formalism will allow us to rigorously describe and analyze its more complex properties.

- Describing mechanisms belonging to physical level (such as partitioning, replications, distributed databases, column-store vs. row-stores and others) as integral part of the model. These functions are implemented by identities and the idea is that abstract identities from the virtual address space can be mapped to identities of the physical address space.

- Better integration with concept-oriented programming by adding data modeling feature to COP and programming mechanisms to COM.

Khoshafian, S.N., & Copeland, G.P. (1986). Object identity. In *Proceedings of the conference on Object-oriented programming systems, languages, and applications (OOPSLA'86)*, *ACM SIGPLAN Notices*, **21**(11), 406–416.

Kuper, G.M., & Vardi, M.Y. (1984). A New Approach to Database Logic. In *Proceedings of the 3rd ACM SIGACT-SIGMOD symposium on Principles of database systems (PODS)* (pp. 86–96).

Kuper, G.M., & Vardi, M.Y. (1993). The Logical Data Model. *ACM Transactions on Database Systems*, **18**(3), 379–413.

Li, C., & Wang, X.S. (1996). A data model for supporting on-line analytical processing. In *Proc. Conference on Information and Knowledge Management* (pp. 81–88), Baltimore, MD.

Maier, D., Ullman, J.D., & Vardi, M.Y. (1984). On the foundation of the universal relation model. *ACM Transactions on Database Systems (TODS)*, **9**(2), 283–308.

Malinowski, E., & Zimanyi, E. (2006). Hierarchies in a multidimensional model: from conceptual modeling to logical representation. *Data & Knowledge Engineering*, **59**(2), 348–377.

Nguyen, T.B., Tjoa, A.M., & Wagner, R.R. (2000). An Object Oriented Multidimensional Data Model for OLAP. *Proc. 1st International Conference on Web-Age Information Management (WAIM'00)*, Shanghai, China.

Peckham, J., & Maryanski, F. (1988). Semantic data models. *ACM Computing Surveys (CSUR)*, **20**(3), 153–189.

Potter, W.D., & Trueblood, R.P. (1988). Traditional, semantic, and hypersemantic approaches to data modeling. *Computer*, **21**(6), 53–63.

Roth, M.A., Korth, H.F., & Silberschatz, A. (1988). Extended Algebra and Calculus for Nested Relational Databases. *ACM Transactions on Database Systems (TODS)*, **13**(4), 389–417.

Savinov, A. (2005a). Hierarchical Multidimensional Modeling in the Concept-Oriented Data Model. *3rd Intl. Conference on Concept Lattices and Their Applications (CLA'05)* (pp. 123–134), Olomouc, Czech Republic.

Savinov, A. (2005b). Concept as a Generalization of Class and Principles of the Concept-Oriented Programming. *Computer Science Journal of Moldova*, **13**(3), 292–335.

Savinov, A. (2006a). Grouping and Aggregation in the Concept-Oriented Data Model. In *Proc. 21st Annual ACM Symposium on Applied Computing (SAC'06)* (pp. 482–486). Dijon, France.

Savinov, A. (2006b). Query by Constraint Propagation in the Concept-Oriented Data Model. *Computer Science Journal of Moldova*, **14**(2), 219–238.